\documentclass[conference]{IEEEtran}
\IEEEoverridecommandlockouts
\usepackage{graphicx}
\usepackage{amsmath,amssymb,amsfonts}
\usepackage{array}
\usepackage[T1]{fontenc}
\usepackage{newtxtext,newtxmath}
\newcommand{\mb}{\mathbf}

\usepackage{stfloats}
\usepackage{diagbox}
\usepackage{graphicx}
\usepackage{footnote}
\usepackage{booktabs}
\usepackage{array}
\usepackage{algorithm}
\usepackage{subeqnarray}
\usepackage{cases}
\usepackage{threeparttable}
\usepackage{color}
\usepackage{epstopdf}
\usepackage{algpseudocode}
\usepackage{bm}
\usepackage{multirow}
\usepackage[labelformat=simple]{subcaption}
\usepackage[font=small]{caption}
\usepackage{adjustbox}
\usepackage[subpreambles=true]{standalone}
\usepackage[xindy, toc, numberedsection, acronym]{glossaries}
\usepackage[hidelinks]{hyperref}
\newacronym{2d}{2D}{two-dimensional}
\newacronym{3d}{3D}{three-dimensional}
\newacronym{3gpp}{3GPP}{3rd generation partnership project}
\newacronym{3msk}{3MSK}{3-level minimum-shift-keying}
\newacronym{4g}{4G}{fourth-generation}
\newacronym{5g}{5G}{fifth-generation}
\newacronym{6dof}{6DoF}{six degrees of freedom}
\newacronym{6g}{6G}{sixth-generation}
\newacronym{abp}{ABP}{authentication by personalisation}
\newacronym{ac}{AC}{alternating current}
\newacronym{aci}{ACI}{adjacent channel interference}
\newacronym{aclr}{ACLR}{adjacent channel leakage ratio}
\newacronym{acpr}{ACPR}{adjacent channel power ratio}
\newacronym{adc}{ADC}{analog-to-digital converter}
\newacronym{adr}{ADR}{adaptive data rate}
\newacronym{aec}{AEC}{aluminum electrolytic capacitor}
\newacronym{aes}{AES}{Advanced Encryption Standard}
\newacronym{afe}{AFE}{analog front end}
\newacronym{ag}{AG}{array gain}
\newacronym{agc}{AGC}{automatic gain controller}
\newacronym{agps}{A-GPS}{Assisted Global Positioning System} %bestaat al :) % Aha, dan maak ik er wel Assisted gps van!
\newacronym{ai}{AI}{artificial intelligence}
\newacronym{alk}{ALK}{Alkaline}
\newacronym{am}{AM}{amplitude modulation}
\newacronym{amam}{AM/AM}{amplitude modulation to amplitude modulation}
\newacronym{amc}{AMC}{automatic modulation classification}
\newacronym{amp}{AMP}{approximate message passing}
\newacronym{ampm}{AM/PM}{amplitude modulation to phase modulation}
\newacronym{aoa}{AOA}{angle-of-arrival}
\newacronym{aod}{AOD}{angle-of-departure}
\newacronym{ap}{AP}{access point}
\newacronym{api}{API}{application program interface}
\newacronym{apt}{APT}{acoustic power transfer}
\newacronym{apu}{APU}{access point unit}
\newacronym{ar}{AR}{augmented reality}
\newacronym{arima}{ARIMA}{Auto-Regressive Integrated Moving Average}
\newacronym{arp}{ARP}{Antenna Reference Point}
\newacronym{asic}{ASIC}{application specific integrated circuit}
\newacronym{ask}{ASK}{amplitude-shift keying}
\newacronym{at}{AT}{ATtention}
\newacronym{auv}{AUV}{autonomous underwater vehicle}
\newacronym{awg}{AWG}{arbitrary waveform generator}
\newacronym{awgn}{AWGN}{additive white Gaussian noise}
\newacronym{baw}{BAW}{bulk acoustic wave}
\newacronym{bb}{BB}{base-band}
\newacronym{bc}{BC}{backscatter communication}
\newacronym{bcjr}{BCJR}{Bahl-Cocke-Jelinek-Raviv}
\newacronym{bd}{BD}{backscatter device}
\newacronym{be}{BE}{Belgium}
\newacronym{ber}{BER}{bit error rate}
\newacronym{bf}{BF}{beamforming}
\newacronym{bga}{BGA}{ball grid array}
\newacronym{bldc}{BLDC}{brushless DC}
\newacronym{bler}{BLER}{block error rate}
\newacronym{bod}{BOD}{Brown-Out Detection}
\newacronym{bom}{BOM}{bill of materials}
\newacronym{bpsk}{BPSK}{binary phase-shift keying}
\newacronym{brp}{BRP}{beam refinement process}
\newacronym{bs}{BS}{base station}
\newacronym{bu}{BU}{booster unit}
\newacronym{bw}{BW}{bandwidth}
\newacronym{ca}{CA}{carrier emitter}
\newacronym{cad}{CAD}{channel activity detection}
\newacronym{cars}{CARS}{calibration reference signal}
\newacronym{cb}{CB}{conjugate beamforming}
\newacronym{cbm}{CBM}{condition based maintenance}
\newacronym{cc}{CC}{constant current}
\newacronym{ccdf}{CCDF}{complementary cumulative distribution function}
\newacronym{ccnn}{CCNN}{circular convolutional neural network}
\newacronym{ccs}{CCS}{correlative channel sounder}
\newacronym{cdf}{CDF}{cumulative distribution function}
\newacronym{cdma}{CDMA}{code division-multiple access}
\newacronym{cdrx}{CDRX}{connected mode DRX}
\newacronym{ce}{CE}{coverage enhancement}
\newacronym{ced}{CED}{cumulative energy density}
\newacronym{ceofdm}{CE-OFDM}{constant-envelope OFDM}
\newacronym{cf}{CF}{cell-free}
\newacronym{cfmmimo}{CF-mMIMO}{cell-free massive MIMO}
\newacronym{cfo}{CFO}{carrier frequency offset}
\newacronym{cir}{CIR}{channel impulse response}
\newacronym{cla}{CLA}{closed-loop approach}
\newacronym{clk}{CLK}{clock}
\newacronym{cmos}{CMOS}{complementary metal oxide semiconductor}
\newacronym{cnn}{CNN}{convolutional neural network}
\newacronym{co}{CO}{combinatorial optimization}
\newacronym{cordic}{CORDIC}{coordinate rotation digital computer}
\newacronym{cost}{COST}{commercial off-the-shelf}
\newacronym{cots}{COTS}{commercial off-the-shelf}
\newacronym{cp}{CP}{cyclic prefix}
\newacronym{cpe}{CPE}{common phase error}
\newacronym{cpfsk}{CPFSK}{continuous phase frequency shift keying}
\newacronym{cpm}{CPM}{continuous phase modulation}
\newacronym{cpt}{CPT}{capacitive power transfer}
\newacronym{cpu}{CPU}{central-processing unit}
\newacronym{cpw}{CPW}{coplanar waveguide}
\newacronym{cqi}{CQI}{channel quality indicator}
\newacronym{cr}{CR}{coding rate}
\newacronym{crc}{CRC}{cyclic redundancy check}
\newacronym{crlb}{CRLB}{Cram\'er-Rao lower bound}
\newacronym{crs}{CRS}{cell reference signal}
\newacronym{cs}{CS}{compressed sensing}
\newacronym{cse}{CSE}{channel state estimation}
\newacronym{csi}{CSI}{channel state information}
\newacronym{csp}{CSP}{contact service point}
\newacronym{css}{CSS}{chirp spread spectrum}
\newacronym{cu}{CU}{central unit}
\newacronym{cv}{CV}{constant voltage}
\newacronym{cw}{CW}{continuous wave}
\newacronym{d2d}{D2D}{device-to-device}
\newacronym{dab}{DAB}{Digital Audio Broadcasting}
\newacronym{dac}{DAC}{digital-to-analog converter}
\newacronym{daq}{DAQ}{data acquisition system}
\newacronym{das}{DAS}{distributed antenna systems}
\newacronym{dbpsk}{DBPSK}{Differential Binary Phase Shift Keying}
\newacronym{dc}{DC}{direct current}
\newacronym{dcc}{DCC}{dynamic cooperation clustering}
\newacronym{ddc}{DDC}{digital down conversion}
\newacronym{de}{DE}{drain efficiency}
\newacronym{dft}{DFT}{discrete Fourier transform}
\newacronym{dftsofdm}{DFT-s-OFDM}{discrete Fourier Transform spread OFDM}
\newacronym{dftsofdmfdss}{DFT-s-OFDM-FDSS}{DFT-s-OFDM with frequency-domain spectral shaping}
\newacronym{dftsofdmfdssse}{DFT-s-OFDM-FDSS-SE}{DFT-s-OFDM with FDSS and spectral extension}
\newacronym{dl}{DL}{downlink}
\newacronym{dlc}{DLC}{Distributed Laser Charging}
\newacronym{dli}{DLI}{direct link interference}
\newacronym{dlt}{DLT}{Distributed Ledger Technology}
\newacronym{dm}{DM}{diffuse multipath}
\newacronym{dma}{DMA}{Direct Memory Access}
\newacronym{dmac}{DMAC}{Direct Memory Access Controller}
\newacronym{dmc}{DMC}{diffuse multipath component}
\newacronym{dmimo}{D-MIMO}{distributed MIMO}
\newacronym{dnn}{DNN}{deep neural network}
\newacronym{doa}{DOA}{direction-of-arrival}
\newacronym{dp}{DP}{differencial privacy}
\newacronym{dpd}{DPD}{digital pre-distortion}
\newacronym{dpdk}{DPDK}{Data Plane Development Kit}
\newacronym{dram}{DRAM}{dynamic random-access memory}
\newacronym{drcs}{$\Delta$RCS}{differential-radar cross section}
\newacronym{drx}{DRX}{Discontinuous Reception Mode}
\newacronym{dsb}{DSB}{double-sideband}
\newacronym{dsl}{DSL}{digital subscriber line}
\newacronym{dsp}{DSP}{digital signal processing}
\newacronym{dss}{DSS}{Dataset Storage Standard}
\newacronym{du}{DU}{Distributed Unit}
\newacronym{duc}{DUC}{digital up-converter}
\newacronym{dvb}{DVB}{Digital Video Broadcasting}
\newacronym{e2e}{E2E}{end-to-end}
\newacronym{easa}{EASA}{European Union Aviation Safety Agency}
\newacronym{ebg}{EBG}{electromagnetic bandgap}
\newacronym{ebw}{EBW}{excess bandwidth}
\newacronym{ec}{EC}{European Commission}
\newacronym{ecc}{ECC}{elliptic curve cryptography}
\newacronym{ecdf}{eCDF}{empirical cumulative distribution function}
\newacronym{ecdlp}{ECDLP}{elliptic curve discrete logarithm problem}
\newacronym{ecsp}{ECSP}{edge computing service point}
\newacronym{edlc}{EDLC}{electrostatic double-layer capacitors}
\newacronym{edrx}{eDRX}{Extended Discontinuous Reception Mode}
\newacronym{ee}{EE}{energy efficiency}
\newacronym{egprs}{EGPRS}{Enhanced Data Rates for GSM Evolution}
\newacronym{eh}{EH}{energy harvesting}
\newacronym{eirp}{EIRP}{equivalent isotropically radiated power}
\newacronym{em}{EM}{electromagnetic}
\newacronym{embb}{eMBB}{enhanced Mobile Broadband}
\newacronym{en}{EN}{energy neutral}
\newacronym{end}{END}{energy-neutral device}
\newacronym{enob}{ENOB}{effective number of bits}
\newacronym{eol}{EoL}{end of life}
\newacronym{ep}{EP}{energy profiler}
\newacronym{epd}{EPD}{electronic paper display}
\newacronym{epu}{EPU}{edge processing unit}
\newacronym{er}{ER}{Energy Receiver}
\newacronym{erp}{ERP}{effective radiation power}
\newacronym[plural=ESCs,firstplural=electronic speed controllers (ESCs)]{esc}{ESC}{electronic speed control}
\newacronym{esd}{ESD}{electrostatic discharge}
\newacronym{esl}{ESL}{electronic shelf label}
\newacronym{esr}{ESR}{equivalent series resistance}
\newacronym{et}{ET}{Energy Transmitter}
\newacronym{etsi}{ETSI}{European Telecommunications Standards Institute}
\newacronym{evd}{EVD}{eigenvalue decomposition}
\newacronym{evm}{EVM}{Error Vector Magnitude}
\newacronym{ewlb}{eWLB}{Embedded Wafer Level Ball Grid Array}
\newacronym{fa}{FA}{federation anchor}
\newacronym{fair}{FAIR}{Findability, Accessibility, Interoperability, and Reuse of digital assets}
\newacronym{fcf}{FCF}{frequency correlation function}
\newacronym{fd}{FD}{front-haul distance}
\newacronym{fdd}{FDD}{frequency-division duplexing}
\newacronym{fde}{FDE}{frequency domain equalizer}
\newacronym{fdm}{FDM}{frequency-division multiplexing}
\newacronym{fdma}{FDMA}{frequency division multiple access}
\newacronym{fdss}{FDSS}{frequency-domain spectral shaping}
\newacronym{fem}{FEM}{finite element analysis}
\newacronym{fembb}{feMBB}{further enhanced mobile broadband}
\newacronym{fft}{FFT}{fast Fourier transform}
\newacronym{fh}{FH}{fronthaul}
\newacronym{fhss}{FHSS}{frequency hopping spread spectrum}
\newacronym{fifo}{FIFO}{First In, First Out}
\newacronym{fim}{FIM}{Fisher information matrix}
\newacronym{fits}{FITS}{Flexible Image Transport System}
\newacronym{fl}{FL}{federated learning}
\newacronym{fom}{FoM}{figure of merit}
\newacronym{fov}{FOV}{field of view}
\newacronym{fpga}{FPGA}{field-programmable gate array}
\newacronym{fr1}{FR1}{frequency range 1}
\newacronym{fr2}{FR2}{frequency range 2}
\newacronym{fram}{FRAM}{Ferroelectric Random Access Memory}
\newacronym{fsk}{FSK}{frequency shift keying}
\newacronym{fspl}{FSPL}{free space path loss}
\newacronym{fss}{FSS}{frequency selective surface}
\newacronym{gb}{GB}{grant-based}
\newacronym{gdpr}{GDPR}{general data protection regulation}
\newacronym{gf}{GF}{grant-free}
\newacronym{gmsk}{GMSK}{Gaussian minimum-shift keying}
\newacronym{gnb}{gNB}{Next Generation Node B}
\newacronym{gni}{GNI}{gross national income}
\newacronym{gnn}{GNN}{graph neural network}
\newacronym{gnss}{GNSS}{global navigation satellite system}
\newacronym{gpclk}{GPCLK}{general purpose clock}
\newacronym{gpio}{GPIO}{General-Purpose Input/Output}
\newacronym{gpl}{GPL}{GNU General Public License}
\newacronym{gprs}{GPRS}{General Packet Radio Services}
\newacronym{gps}{GPS}{Global Positioning System}
\newacronym{gpu}{GPU}{graphical processing unit}
\newacronym{grc}{GRC}{GNU Radio Companion}
\newacronym{gscm}{GSCM}{geometry‐based stochastic model}
\newacronym{gsm}{GSM}{Global System for Mobile Communications}  %
\newacronym{gspm}{GSpM}{Generalized spatial modulation}
\newacronym{gwp}{GWP}{Global Warming Potential}
\newacronym{harq}{HARQ}{hybrid automatic repeat request}
\newacronym{hat}{HAT}{hardware attached on top}
\newacronym{hcs}{HCS}{human-centric services}
\newacronym{hdf5}{HDF5}{Hierarchical Data Format version 5}
\newacronym{hfss}{HFSS}{High Frequency Simulator Software}
\newacronym{hmd}{HMD}{head-mounted display}
\newacronym{hpbm}{HPBM}{half power beam width}
\newacronym{HyMPRo}{HyMPRo}{Hybrid Multi-Path Routing algorithm}
\newacronym{i.i.d.}{i.i.d.}{independent and identically distributed}
\newacronym{i2c}{I2C}{Inter-Integrated Circuit}
\newacronym{iaq}{IAQ}{Indoor Air Quality}
\newacronym{ib}{IB}{in-band}
\newacronym{ibbc}{IBBC}{inter-band beam configuration}
\newacronym{ibo}{IBO}{input back-off}
\newacronym{ic}{IC}{integrated circuit}
\newacronym{iccs}{ICCS}{Ilmsens correlative channel sounder}
\newacronym{ici}{ICI}{intercarrier interference}
\newacronym{icnirp}{ICNIRP}{International Commission on Non-Ionizing Radiation Protection}
\newacronym{id}{ID}{information decoding}
\newacronym{idft}{IDFT}{inverse discrete Fourier transform}
\newacronym{idxm}{IDXM}{index modulation}
\newacronym{if}{IF}{intermediate-frequency}
\newacronym{ifft}{IFFT}{inverse fast-Fourier-transform}
\newacronym{iid}{i.i.d.}{independently and identically distributed}
\newacronym{iis}{IIS}{integrated information system}
\newacronym{im}{IM}{intermodulation}
\newacronym{imd}{IMD}{intermodulation distortion}
\newacronym{imu}{IMU}{inertial measurement unit}
\newacronym{inh}{InH}{indoor hotspot office}
\newacronym{io}{IO}{input/output}
\newacronym{ioe}{IoE}{Internet of Everything}
\newacronym{iot}{IoT}{Internet of Things}
\newacronym{ipt}{IPT}{inductive power transfer}
\newacronym{ipy}{IPY}{Interventions per Year}
\newacronym{iq}{IQ}{in-phase and quadrature}
\newacronym{iqi}{IQI}{IQ imbalance}
\newacronym{ir}{IR}{infrared}
\newacronym{isi}{ISI}{intersymbol interference}
\newacronym{ism}{ISM}{industrial, scientific and medical}
\newacronym{isp}{ISP}{internet service provider}
\newacronym{itu}{ITU}{International Telecommunication Union}
\newacronym{jesd}{JESD}{Joint Electron Devices Engineering Council}
\newacronym{jfet}{JFET}{junction field effect transistor}
\newacronym{kpi}{KPI}{key performance indicator}
\newacronym{ktofdm}{KT-DFT-s-OFDM}{known-tail-DFT-s-OFDM}
\newacronym{kvi}{KVI}{key value indicator}
\newacronym{larva}{LARVA}{LARge Virtual Array}
\newacronym{lca}{LCA}{life cycle assessment}
\newacronym{lco}{LCO}{lithium cobalt oxide}
\newacronym{ldo}{LDO}{Low-dropout voltage regulator}
\newacronym{ldpc}{LDPC}{low-density parity-check}
\newacronym{led}{LED}{Light Emitting Diode}
\newacronym{less}{LESS}{Low Energy Scheduler Solution}
\newacronym{lfp}{LFP}{lithium iron phosphate}
\newacronym{lib}{LIB}{Lithium-Ion Battery}
\newacronym{lic}{LIC}{lithium-ion capacitor}
\newacronym{lid}{LID}{Lithium Iron Disulfide}
\newacronym{lidar}{LiDAR}{light detection and ranging}
\newacronym{liion}{Li-ion}{lithium-ion}
\newacronym{lipo}{LiPo}{lithium polymer}
\newacronym{lis}{LIS}{large intelligent surface}
\newacronym{llh}{LLH}{log-likelihood}
\newacronym{lls}{LLS}{link-level simulation}
\newacronym{lmd}{LMD}{Lithium Manganese Dioxide}
\newacronym{lmmse}{LMMSE}{least minimum mean square error}
\newacronym{lmo}{LMO}{lithium ion manganese oxide}
\newacronym{lna}{LNA}{low-noise amplifier}
\newacronym{lo}{LO}{local oscillator}
\newacronym{lora}{LoRa}{long range}
\newacronym{lorawan}{LoRaWAN}{long-range wide-area network}
\newacronym{los}{LoS}{line-of-sight}
\newacronym{lp}{LP}{linear programming}
\newacronym{lpf}{LPF}{low-pass filter}
\newacronym{lpt}{LPT}{laser power transfer}
\newacronym{lpwa}{LPWA}{Low Power Wide Area}
\newacronym{lpwan}{LPWAN}{low-power wide-area network}
\newacronym{lpwans}{LPWANs}{Low-Power Wide-Area Networks}
\newacronym{lqi}{LQI}{link quality indicator}
\newacronym{lrelu}{LReLU}{leaky rectified linear unit}
\newacronym{lrt}{LRT}{likelihood-ratio test}
\newacronym{ls}{LS}{least squares}
\newacronym{lsa}{LSA}{large synthetic array}
\newacronym{lsf}{LSF}{large-scale fading}
\newacronym{lsfc}{LSFC}{large-scale fading component}
\newacronym{lstm}{LSTM}{Long Short-Term Memory}
\newacronym{ltc}{LTC}{lithium thionyl chloride}
\newacronym{lte}{LTE}{Long Term Evolution}
\newacronym{lti}{LTI}{linear time-invariant}
\newacronym{lto}{LTO}{lithium titanate}
\newacronym{lusta}{LUSTA 5G }{Logistique mUltimodale Sécuritaire Téléopérée \& Autonome 5G}
\newacronym{m2m}{M2M}{machine to machine}
\newacronym{mac}{MAC}{Medium Access Control}
\newacronym{mate}{MATE}{millimeter-wave MIMO testbed}
\newacronym{mc}{MC}{Monte Carlo}
\newacronym{mcl}{MCL}{Maximum Coupling Loss}
\newacronym{mcs}{MCS}{modulation and coding scheme}
\newacronym{mcu}{MCU}{microcontroller unit}
\newacronym{mec}{MEC}{multi-access edge computing}
\newacronym{mems}{MEMS}{micro-electromechanical systems}
\newacronym{mf}{MF}{matched filter}
\newacronym{mimo}{MIMO}{multiple-input multiple-output}
\newacronym{miso}{MISO}{multiple-input single-output}
\newacronym{ml}{ML}{machine learning}
\newacronym{mle}{MLE}{maximum likelihood estimation}
\newacronym{mlp}{MLP}{multilayer perceptron}
\newacronym{mmic}{MMIC}{monolithic microwave integrated circuit}
\newacronym{mmimo}{mMIMO}{massive MIMO}
\newacronym{mmse}{MMSE}{minimum mean square error}
\newacronym{mmtc}{mMTC}{massive machine-typed communication}
\newacronym{mmwave}{mmWave}{millimeter wave}
\newacronym{mn}{MN}{matching network}
\newacronym{mosfet}{MOSFET}{metal-oxide semiconductor field effect transistor}
\newacronym{mpc}{MPC}{multipath component}
\newacronym{mppt}{MPPT}{maximum power point tracking}
\newacronym{mr}{MR}{maximum ratio}
\newacronym{mrc}{MRC}{maximum ratio combining}
\newacronym{mrc_em}{MRC}{maximum ratio combining}
\newacronym{mrc_EM}{MRC}{Magnetic Resonance Coupling}
\newacronym{mrt}{MRT}{maximum ratio transmission}
\newacronym{mse}{MSE}{mean square error}
\newacronym{msk}{MSK}{Minimum-Shift Keying}
\newacronym{mtc}{MTC}{Machine-Type Communication}
\newacronym{multi-rat}{Multi-RAT}{multiple radio access technology}
\newacronym{multirat}{Multi-RAT}{Multiple Radio Access Technology}
\newacronym{music}{MUSIC}{MUltiple SIgnal Classification}
\newacronym{navauwall}{NAVAUWALL}{AUtomated NAVigation in WALLonia}
\newacronym{nb}{NB}{narrowband}
\newacronym{nbiot}{NB-IoT}{narrowband IoT}
\newacronym{nca}{NCA}{nickel cobalt aluminum}
\newacronym{netcdf}{NetCDF}{Network Common Data Form}
\newacronym{nf}{NF}{noise figure}
\newacronym{nfc}{NFC}{near-field communication}
\newacronym{nfv}{NFV}{network function virtualization}
\newacronym{ngmn}{NGMN}{Next Generation Mobile Networks }
\newacronym{ni}{NI}{National Instruments}
\newacronym{nicd}{NiCd}{nikkel cadmium}
\newacronym{nimh}{NiMH}{nikkel metal hydride}
\newacronym{nlos}{NLoS}{non-line-of-sight}
\newacronym{nmc}{NMC}{nickel manganese cobalt}
\newacronym{nmos}{nMOS}{n-channel metal-oxide semiconductor}
\newacronym{nn}{NN}{neural network}
\newacronym{nnls}{NNLS}{non-negative least squares}
\newacronym{noma}{NOMA}{non-orthogonal multiple access}
\newacronym{np}{NP}{Neyman-Pearson}
\newacronym{npbch}{NPBCH}{Narrowband Physical Broadcast Channel}
\newacronym{npss}{NPSS}{Narrow Band Primay Synchronization Signal}
\newacronym{nr}{NR}{New Radio}
\newacronym{nrs}{NRS}{Narrow Band Reference Signal}
\newacronym{nsss}{NSSS}{Narrowband Secondary Synchronization Signal}
\newacronym{ntp}{NTP}{network time protocol}
\newacronym{oai}{OAI}{OpenAirInterface} % no spelling mistake, it is spelled all glued to eachother...
\newacronym{obw}{OBW}{occupied bandwidth}
\newacronym{ofdm}{OFDM}{orthogonal frequency-division multiplexing}
\newacronym{ofdma}{OFDMA}{orthogonal frequency-division multiple access}
\newacronym{ofdmim}{OFDM-IM}{OFDM with index modulation}
\newacronym{olos}{OLoS}{obstructed-line-of-sight}
\newacronym{oma}{OMA}{orthogonal multiple access}
\newacronym{oob}{OOB}{out-of-band}
\newacronym{ook}{OOK}{on-off keying}
\newacronym{opbo}{OPBO}{output power backoff}
\newacronym{oran}{O-RAN}{open radio-access network}
\newacronym{os}{OS}{operating system}
\newacronym{ota}{OTA}{over-the-air}
\newacronym{otaa}{OTAA}{over-the-air authentication}
\newacronym{p1}{P1}{Phase 1}
\newacronym{p2}{P2}{Phase 2}
\newacronym{p2p}{P2P}{point-to-point}
\newacronym{pa}{PA}{power amplifier}
\newacronym{pae}{PAE}{power-added efficiency}
\newacronym{pam}{PAM}{pulse amplitude modulation}
\newacronym{pana}{PanA}{Panel A}
\newacronym{panb}{PanB}{Panel B}
\newacronym{papr}{PAPR}{peak-to-average power ratio}
\newacronym{pc}{PC}{pilot count}
\newacronym{pcb}{PCB}{printed circuit board}
\newacronym{pcg}{PCG}{power consumption gain}
\newacronym{pcie}{PCIe}{Peripheral Component Interconnect Express}
\newacronym{pcsi}{PCSI}{perfect channel state information}
\newacronym{pd}{PD}{powered device}
\newacronym{pdcch}{PDCCH}{physical downlink control channel}
\newacronym{pdf}{PDF}{probability density function}
\newacronym{pdp}{PDP}{power delay profile}
\newacronym{pdsch}{PDSCH}{physical downlink shared channel}
\newacronym{pe}{PE}{processing element}
\newacronym{peb}{PEB}{positioning error bound}
\newacronym{per}{PER}{packet error rate}
\newacronym{pet}{PET}{privacy enhancing technology}
\newacronym{pg}{PG}{path gain}
\newacronym{pgd}{PGD}{proximal gradient descent}
\newacronym{phy}{PHY}{physical}
\newacronym{pki}{PKI}{public key infrastucture}
\newacronym{pl}{PL}{path loss}
\newacronym{pla}{PLA}{physically large array}
\newacronym{pll}{PLL}{phase-locked loop}
\newacronym[plural=PMs,firstplural=person months (PMs)]{pm}{PM}{person month}
\newacronym{pmf}{PMF}{polymer microwave fiber}
\newacronym{pmu}{PMU}{power management unit}
\newacronym{pn}{PN}{pseudo-noise}
\newacronym{po}{PO}{phase offset}
\newacronym{poc}{PoC}{proof of concept}
\newacronym{poe}{PoE}{power-over-Ethernet}
\newacronym{pps}{1PPS}{pulse per second}
\newacronym{pr}{PR}{phase reversal}
\newacronym{prb}{PRB}{Physical Resource Block}
\newacronym{prbs}{PRBs}{Physical Resource Blocks}
\newacronym{prs}{PRS}{Peripheral Reflex System}
\newacronym{ps}{PS}{Processing System}
\newacronym{psd}{PSD}{power spectral density}
\newacronym{pse}{PSE}{power sourcing equipment}
\newacronym{psk}{PSK}{phase shift keying}
\newacronym{psm}{PSM}{power saving mode}
\newacronym{pss}{PSS}{primary synchronisation signal}
\newacronym{ptp}{PTP}{precision-time protocol}
\newacronym{ptrs}{PTRS}{Phase-Tracking Reference Signals}
\newacronym{ptw}{PTW}{paging time window}
\newacronym{pv}{PV}{photovoltaic}
\newacronym{pw}{PW}{planar wavefront}
\newacronym{pwm}{PWM}{pulse width modulation}
\newacronym{qam}{QAM}{quadrature amplitude modulation}
\newacronym{qos}{QoS}{quality-of-service}
\newacronym{qpsk}{QPSK}{quadrature phase-shift keying}
\newacronym{qrrls}{QR-RLS}{QR decomposition based recursive least squares}
\newacronym{quadriga}{QuaDRiGa}{QUAsi Deterministic RadIo channel GenerAtor}
\newacronym{ra}{RA}{Random Access}
\newacronym{ram}{RAM}{random-access memory}
\newacronym{ran}{RAN}{radio access network}
\newacronym{rps}{RPS}{Random-Phase Sweeping}
\newacronym{rar}{RAR}{Random Access Response}
\newacronym{rat}{RAT}{radio access technology}
\newacronym{rb}{Rb}{Rubidium}
\newacronym{rbs}{RBS}{radio base station}
\newacronym{rbw}{RBW}{resolution bandwidth}
\newacronym{rc}{RC}{raised-cosine}
\newacronym{rcs}{RCS}{radar cross section}
\newacronym{rdl}{RDL}{redistribution layer}
\newacronym{re}{RE}{radio element}
\newacronym{relu}{ReLU}{rectified linear unit}
\newacronym{rf}{RF}{radio frequency}
\newacronym{rfeh}{RFEH}{radio frequency energy harvesting}
\newacronym{rfic}{RFIC}{radio-frequency integrated circuit}
\newacronym{rfid}{RFID}{radio frequency identification}
\newacronym{rfpt}{RFPT}{radio frequency power transfer}
\newacronym{rfsoc}{RFSoC}{Radio Frequency System-on-Chip}
\newacronym{rir}{RIR}{room impulse response}
\newacronym{ris}{RIS}{reflective intelligent surface}%reconfigurable?
\newacronym{rllmtc}{RLLMTC}{reliable low latency machine type communication}
\newacronym{rls}{RLS}{recursive least squares}
\newacronym{rms}{RMS}{root-mean-square}
\newacronym{rmse}{RMSE}{root-mean-square error}
\newacronym{rmt}{RMT}{random matrix theory}
\newacronym{rof}{RoF}{radio-over-fiber}
\newacronym{ros}{ROS}{robot operating system}
\newacronym{rpi}{RPi}{Raspberry Pi}
\newacronym{rrc}{RRC}{Radio Resource Connection}
\newacronym{rreq}{RREQ}{route request packet}
\newacronym{rsrp}{RSRP}{Reference Signals Received Power}
\newacronym{rsrq}{RSRQ}{Reference Signal Received Quality}
\newacronym{rss}{RSS}{received signal strength}
\newacronym{rssi}{RSSI}{received signal strength indicator}
\newacronym{rtc}{RTC}{real time clock}
\newacronym{rtf}{RTF}{reader talks first}
\newacronym{rtk}{RTK}{real time kinematics}
\newacronym{rts}{RTS}{ray tracing simulator}
\newacronym{ru}{RU}{Radio Unit}
\newacronym{rv}{RV}{random variable}
\newacronym{rw}{RW}{RadioWeaves}
\newacronym{rx}{RX}{receiver}
\newacronym{rzf}{RZF}{regularized zero forcing}
\newacronym{s-parameter}{S-parameter}{scattering parameter}
\newacronym{sa}{SA}{synchronization anchor}
\newacronym{s-a}{SA}{Service \& System Aspects}
\newacronym{sa5}{SA}{Stand Alone}
\newacronym{sar}{SAR}{specific absorption rate}
\newacronym{sbl}{SBL}{sparse Bayesian learning}
\newacronym{sc}{SC}{single carrier}
\newacronym{scfde}{SC-FDE}{single-carrier modulation with frequency-domain-equalization}
\newacronym{scfdma}{SCFDMA}{single-carrier frequency division multiple access}
\newacronym{scs}{SCS}{sub-carrier spacing}
\newacronym{sdg}{SDG}{Sustainable Development Goal}
\newacronym{sdm}{SDM}{sigma-delta modulator}
\newacronym{sdma}{SDMA}{spatial-division multiple access}
\newacronym{sdn}{SDN}{software-defined network}
\newacronym{sdof}{SDoF}{sigma-delta over fiber}
\newacronym{sdr}{SDR}{software-defined radio}
\newacronym{se}{SE}{spectral efficiency}
\newacronym{sei}{SEI}{specific emitter identification}
\newacronym{ser}{SER}{symbol-error rate}
\newacronym{sf}{SF}{spreading factor}
\newacronym{sfn}{SFN}{single frequency network}
\newacronym{sfp}{SFP}{small form-factor pluggable}
\newacronym{sha}{SHA}{Secure Hash Algorithm}
\newacronym{SigMF}{SigMF}{Signal Metadata Format}
\newacronym{simo}{SIMO}{single-input multiple-output}
\newacronym{sinr}{SINR}{signal-to-interference-plus-noise ratio}
\newacronym{siso}{SISO}{single-input single-output}
\newacronym{slam}{SLAM}{simultaneous localization and mapping}
\newacronym{slc}{SLC}{spatial leakage suppression}
\newacronym{slerp}{SLERP}{spherical linear interpolation}
\newacronym{sma}{SMA}{SubMiniature version A}
\newacronym{smc}{SMC}{specular multipath component}
\newacronym{smps}{SMPS}{switched mode power supply}
\newacronym{sndr}{SNDR}{signal-to-noise-and-distortion ratio}
\newacronym{snidr}{SNIDR}{signal-to-noise-and-interference-and-distortion ratio}
\newacronym{snir}{SNIR}{signal-to-interference-plus-noise ratio}
\newacronym{snr}{SNR}{signal-to-noise ratio}
\newacronym{soc}{SoC}{state of charge}
\newacronym{SoC}{SOC}{System on Chip}
\newacronym{sp}{SP}{service point}
\newacronym{spdt}{SPDT}{single pole double throw}
\newacronym{spi}{SPI}{Serial Peripheral Interface}
\newacronym{spst}{SPST}{single pole single throw}
\newacronym{sram}{SRAM}{static random-access memory}
\newacronym{srd}{SRD}{short-range device}
\newacronym{srls}{SRLS}{standard recursive least squares}
\newacronym{srs}{SRS}{Sounding Reference Signal}
\newacronym{ssb}{SSB}{synchronisation signal block}
\newacronym{ssd}{SSD}{solid state drive}
\newacronym{ssq}{SSQ}{simulator sickness questionnaire}
\newacronym{steam}{STEAM}{science, technology, engineering, the arts, and mathematics}
\newacronym{svd}{SVD}{singular value decomposition}
\newacronym{sw}{SW}{spherical wavefront}
\newacronym{swipt}{SWIPT}{simultaneous wireless information and power transfer}
\newacronym{synce}{SyncE}{Synchronous Ethernet}
\newacronym{ta}{TA}{timing advance}
\newacronym{tau}{TAU}{tracking area update}
\newacronym{tcer}{TCER}{transported to consumed energy ratio}
\newacronym{tcp}{TCP}{Transmission Control Protocol}
\newacronym{tcxo}{TCXO}{temperature compensated crystal oscillator}
\newacronym{tdce}{TD-CE}{time-domain compression and expansion}
\newacronym{tdd}{TDD}{time division duplexing}
\newacronym{tdma}{TDMA}{time division-multiple access}
\newacronym{tdoa}{TDOA}{time-difference-of-arrival}
\newacronym{thz}{THz}{Terahertz}
\newacronym{tl}{TL}{transfer learning}
\newacronym{to}{TO}{timing offset}
\newacronym{toa}{TOA}{time-of-arrival}
\newacronym{tof}{ToF}{time-of-flight}
\newacronym{tosm}{TOSM}{through-open-short-match}
\newacronym{tpms}{TPMS}{Tire-Pressure Monitoring System}
\newacronym{trl}{TRL}{technology readyness level}
\newacronym{trp}{TRP}{Transmission Reception Point}
\newacronym{tsn}{TSN}{time-sensitive networking}
\newacronym{ttf}{TTF}{tag talks first}
\newacronym{ttff}{TTFF}{Time To First Fix}
\newacronym{tti}{TTI}{transmission time interval}
\newacronym{ttm}{TTM}{time to market}
\newacronym{ttn}{TTN}{The Things Network}
\newacronym{tx}{TX}{transmitter}
\newacronym{uart}{UART}{Universal Asynchronous Receiver/Transmitter}
\newacronym{uav}{UAV}{unmanned aerial vehicle}
\newacronym{uc}{UC}{use case}
\newacronym{ucie}{UCIe}{Universal Chiplet Interconnect Express}
\newacronym{udp}{UDP}{User Datagram Protocol}
\newacronym{ue}{UE}{user equipment}
\newacronym{ugv}{UGV}{unmanned ground vehicle}
\newacronym{uhd}{UHD}{USRP hardware driver}
\newacronym{uhf}{UHF}{ultra-high frequency}
\newacronym{ul}{UL}{uplink}
\newacronym{ula}{ULA}{uniform linear array}
\newacronym{uMIMO}{$\mu$-MIMO}{ultra-massive Multiple-Input Multiple-Output}
\newacronym{ummtc}{umMTC}{ultra massive machine type communication}
\newacronym{un}{UN}{United Nations}
\newacronym{unow}{uNOW}{unified non-orthogonal waveform}
\newacronym{upa}{UPA}{uniform planar array}
\newacronym{ura}{URA}{uniform rectangular array}
\newacronym{urllc}{URLLC}{ultra-reliable low-latency communications}
\newacronym{usrp}{USRP}{universal software radio peripheral}
\newacronym{uv}{UV}{unmanned vehicle}
\newacronym{uw}{UW}{unique word}
\newacronym{uwb}{UWB}{ultrawideband}
\newacronym{uwofdm}{UW-DFT-s-OFDM}{unique word DFT-s-OFDM}
\newacronym{v2v}{V2V}{vehicle-to-vehicle}
\newacronym{vco}{VCO}{voltage-controlled oscillator}
\newacronym{vep}{VEP}{virtual edge platform}
\newacronym{vlc}{VLC}{visible light communication}
\newacronym{vlp}{VLP}{visible light positioning}
\newacronym{vna}{VNA}{vector network analyzer}
\newacronym{voc}{VOC}{Voltatile Organic Compound}
\newacronym{votable}{VOTable}{Virtual Observatory Table}
\newacronym{vr}{VR}{virtual reality}
\newacronym{vuca}{VUCA}{volatile, uncertain, complex and ambiguous}
\newacronym{wb}{WB}{wideband}
\newacronym{wban}{WBAN}{wireless body area network}
\newacronym{wd}{WD}{Wireless distance}
\newacronym{wimax}{WiMAX}{Worldwide Interoperability for Microwave Access}
\newacronym{wlan}{WLAN}{wireless LAN}
\newacronym[plural=WPs,firstplural=work packages (WPs)]{wp}{WP}{work package}
\newacronym{wpt}{WPT}{wireless power transfer}
\newacronym{wr}{WR}{White Rabbit}
\newacronym{wrsn}{WRSN}{wireless rechargeable sensor network}
\newacronym{wsn}{WSN}{wireless sensor network}
\newacronym{xets}{XETS}{cross exponentially tapered slot}
\newacronym{xlmimo}{XL-MIMO}{extremely large-scale MIMO}
\newacronym{xr}{XR}{extended reality}
\newacronym{z3ro}{Z3RO}{zero 3\textsuperscript{rd} order distortion}
\newacronym{zf}{ZF}{zero-forcing}
\newacronym{zmcscg}{ZMCSCG}{zero mean circularly symmetric complex Gaussian}
\newacronym{zmq}{ZMQ}{ZeroMQ}
\newacronym{ztofdm}{ZT-DFT-s-OFDM}{zero-tail DFT-s-OFDM}

\usepackage{tikz}
\usepackage{caption,subcaption}
\usetikzlibrary{arrows.meta,positioning,calc}

\usepackage[capitalise]{cleveref}

\usepackage{xifthen}
\newcommand{\prob}[1][]{%requires xifthen package%
\ifthenelse{\isempty{#1}}%
      {\ensuremath{P}}%
    {\ensuremath{P\left\(#1\right\)}}%
}

\usepackage[per-mode=symbol]{siunitx}
\DeclareSIUnit{\dBm}{dBm}
\usepackage[backend=biber, style=ieee, maxnames=1, minnames=1]{biblatex}
\AtBeginBibliography{\footnotesize}

\begin{document}

\title{Testbed Evaluation of AI-based Precoding in Distributed MIMO Systems}

\author{
    \IEEEauthorblockN{Tianzheng Miao$^{*}$, Thomas Feys$^{*}$, Gilles Callebaut$^{*}$, Jarne Van Mulders$^{*}$,\\ Md Arifur Rahman$^{\dagger}$, François Rottenberg$^{*}$}
    \IEEEauthorblockA{
        $^{*}$KU Leuven, Dept. Electrical Engineering (ESAT), B-9000 Ghent, Belgium\\
        $^{\dagger}$Research and Innovation Department, IS-Wireless, Piaseczno, Poland\\
        \{tianzheng.miao, thomas.feys, gilles.callebaut, jarne.vanmulders, francois.rottenberg\}@kuleuven.be,\\ a.rahman@is-wireless.com
    }
    \thanks{This work was supported by the European Union's Horizon 2022 research and innovation program under Grant Agreement No 101120332 (EMPOWER-6G), and by the Research Foundation - Flanders (FWO) through a Junior Postdoctoral Fellowship, project “Cocoon: Towards Fluid Energy-Efficient Open Access Networks through Citizen's Co-Creation” (grant/application no. 12A2V25N).}
    % \thanks{This work has received funding from the European Union's Horizon 2022 research and innovation program under Grant Agreement No 101120332 (EMPOWER-6G). The source code and datasets are available at \url{https://github.com/Agata872/GNN_based_downlink_precoder_Deployment}.}
}

\maketitle
\begin{abstract}
\Gls{dmimo} has emerged as a key architecture for future \gls{6g} networks, enabling cooperative transmission across spatially distributed \glspl{ap}. 
However, most existing studies rely on idealized channel models and lack hardware validation, leaving a gap between algorithmic design and practical deployment. 
Meanwhile, recent advances in \gls{ai}-driven precoding have shown strong potential for learning nonlinear channel-to-precoder mappings, but their real-world deployment remains limited due to challenges in data collection and model generalization. 
This work presents a framework for implementing and validating an \gls{ai}-based precoder on a \gls{dmimo} testbed with hardware reciprocity calibration. 
A pre-trained \gls{gnn}-based model is fine-tuned using real-world \gls{csi} collected from the Techtile platform and evaluated under both interpolation and extrapolation scenarios before end-to-end validation. 
Experimental results demonstrate a \SI{15.7}{\percent} performance gain over the pre-trained model in the multi-user case after fine-tuning, while in the single-user scenario the model achieves near-\gls{mrt} performance with less than \SI{0.7}{bits/channel\,use} degradation out of a total throughput of \SI{5.19}{bits/channel\,use} on unseen positions. 
Further analysis confirms the data efficiency of real-world measurements, showing consistent gains with increasing training samples, and end-to-end validation verifies coherent power focusing comparable to \gls{mrt}.
\end{abstract}

\glsresetall
\section{introduction}
% \printglossary[type=\acronymtype, title=glossary]
% \printglossary[title=glossary]

\Gls{dmimo} has emerged as a key technology for future \gls{6g} systems, featuring numerous distributed \glspl{ap} interconnected with one or more \glspl{cpu}. 
This architecture enables cooperative multi-user transmission through joint signal processing based on users' \gls{csi}, either locally estimated or centrally aggregated~\cite{ref1}. 
However, as the network scales, efficiently managing the growing \gls{csi} volume and maintaining robust precoding across diverse propagation conditions remain major challenges~\cite{ref22}. 
These challenges motivate the development of scalable and generalizable precoding methods that can adapt to real-world impairments beyond idealized assumptions, with testbed validation playing a crucial role in bridging theory and practice.

When dealing with hardware non-idealities, two primary strategies are typically adopted: linearizing the impairment or compensating for it through calibration. 
For the former,~\cite{ref28} proposes an \gls{ota} digital predistortion scheme that jointly mitigates nonlinear distortion and reciprocity mismatch using mutual coupling measurements, eliminating the need for dedicated calibration hardware. 
For the latter,~\cite{ref29} presents an experimental implementation of a \gls{tdd} reciprocity-based self-calibration method on a \gls{sdr} testbed, enabling coherent downlink transmission without external calibration nodes and validating its performance under real hardware impairments.

While reciprocity calibration remains essential for restoring channel reciprocity and mitigating hardware mismatches, most studies have focused on analytical or simulated evaluation with limited real-hardware validation~\cite{ref31}. 
Building on these calibration foundations, \gls{ai}-based approaches have emerged to enhance precoding by learning nonlinear mappings from \gls{csi} to transmission strategies that better adapt to complex propagation~\cite{ref23}. 
Driven by this potential, \gls{ai} has gained increasing attention in \gls{3gpp} as an enabler for intelligent and adaptive operation in 5G-Advanced and \gls{6g}, though deployment remains limited by model generalization and data constraints~\cite{ref21}. 
Our previous work~\cite{ref14} introduced a fine-tuning framework that adapts pre-trained \gls{gnn}-based precoders to measured distributed channels, bridging synthetic and real-world environments. 
These challenges further highlight the need for testbed-based validation under realistic propagation and synchronization conditions.

Motivated by these challenges and our previous work, this study aims to connect simulation-based \gls{ai}-driven precoding studies with practical \gls{dmimo} deployments. 
While most existing works evaluate learning-based precoders under synthetic datasets, few have validated their feasibility on real hardware platforms subject to synchronization errors, reciprocity mismatch, and environmental variability.  
Building upon this motivation, the main contributions of this work are summarized as follows:  
\begin{itemize}
    \item We design and implement a network-side deployment framework that enables the effective integration of the fine-tuned \gls{gnn}-based precoder into a practical \gls{dmimo} system. The framework supports offline training and online inference at the \gls{cpu}, facilitating scalable and coherent multi-antenna transmission.
    \item We evaluate the interpolation and extrapolation performance of the fine-tuned \gls{gnn}-based precoder against conventional benchmarks, and analyze its data efficiency with respect to the number of real-world training samples.  
    \item We conduct \gls{ota} experiments on a calibrated \gls{dmimo} testbed to validate the end-to-end performance of the deployed \gls{gnn}-based precoder, demonstrating its capability to achieve near-coherent power focusing comparable to the benchmark.
\end{itemize}
The rest of this paper is organized as follows. 
\Cref{sec:system model} introduces the system model, including the \gls{dmimo} architecture and the motivation behind the precoding schemes adopted in this work. 
\Cref{sec:system implemetation} details the practical implementation of the \gls{dmimo} testbed, including synchronization procedures, reciprocity calibration, and the integration of the \gls{gnn}-based precoding framework for real-life operation.
In \cref{sec:results}, we evaluate the performance of the fine-tuned \gls{gnn}-based precoder against conventional benchmarks using real-world datasets, and validate its effectiveness through hardware experiments on the testbed. 
Finally, \cref{sec:conclusions} concludes the paper and outlines directions for future research.

\textit{Notations:} Boldface lowercase and uppercase letters denote vectors and matrices.  \((\cdot)^\mathrm{T}\) and \((\cdot)^H\) indicate the transpose and conjugate transpose. The Frobenius norm is given by \(\|\cdot\|_F\). The set of complex numbers is represented by \(\mathbb{C}\).

\section{System Model}
\label{sec:system model}
\subsection{Distributed MIMO Model}
\begin{figure}[htpb]
\centering
\includegraphics[width=0.4\textwidth]{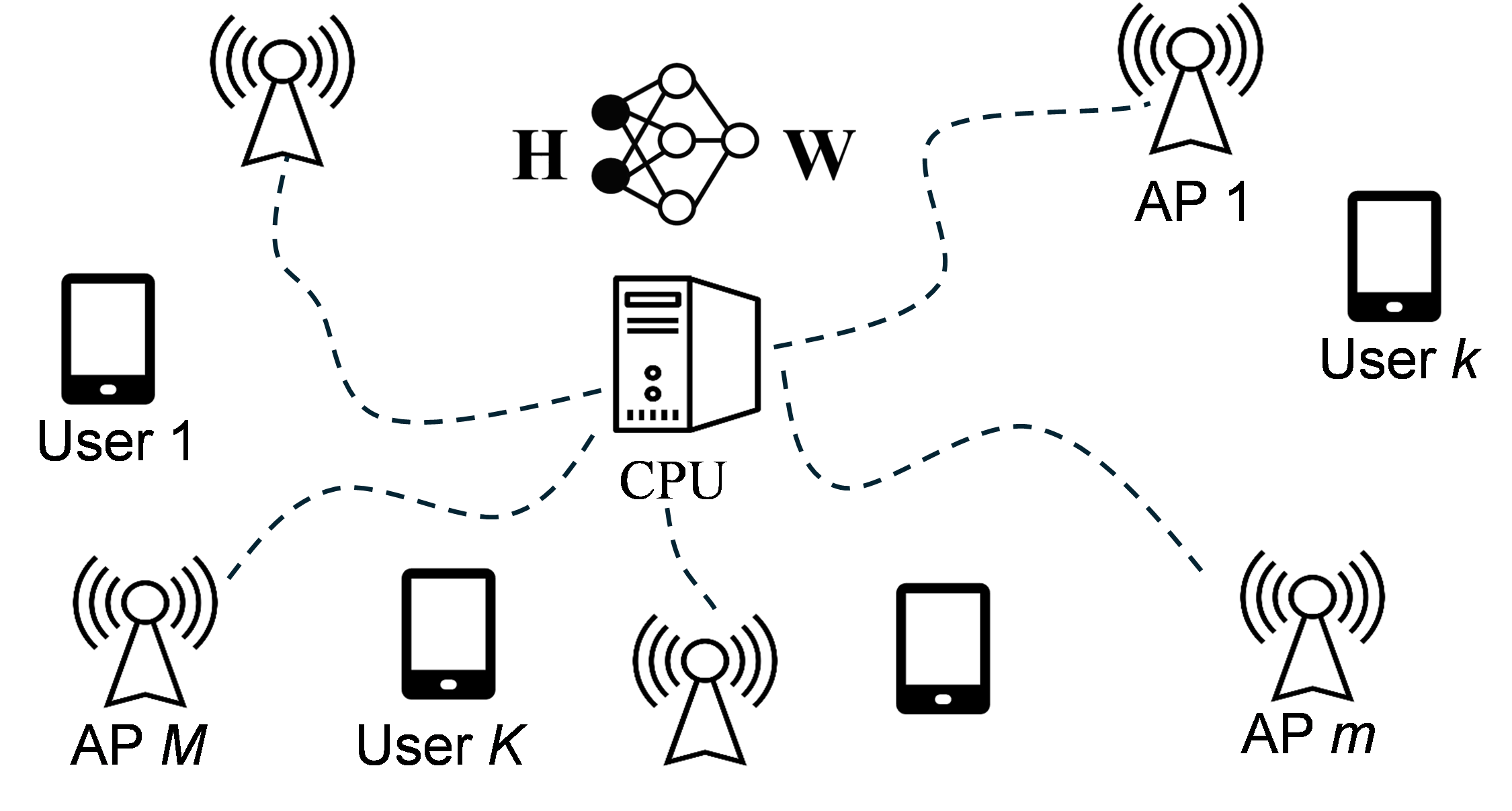}
\caption{System model of the considered \gls{dmimo} network. Distributed \glspl{ap} serve multiple single-antenna \glspl{ue} with coordination by a \gls{cpu}, which handles joint signal processing.}
\label{fig:6}
\end{figure}

In this work, we consider a \gls{dmimo} network consisting of \( M \) single-antenna \glspl{ap} and \( K \) single-antenna \glspl{ue}, as illustrated in~\cref{fig:6}. 
Let \( m = 1, 2, \dots, M \) and \( k = 1, 2, \dots, K \) index the \glspl{ap} and \glspl{ue}, respectively. 
We consider an ideal case where all \glspl{ap} are connected to a central \gls{cpu} via fronthaul links, which can be either wired or wireless in practical \gls{dmimo} deployments. 
In the testbed used in this work, the \glspl{ap}, implemented with \glspl{rpi} as distributed processing units, are connected to the \gls{cpu} through Gigabit Ethernet links, enabling centralized channel estimation and joint precoding based on the global \gls{csi}.

A fully centralized downlink transmission scenario is considered, where the \glspl{ap} are randomly distributed within a given area, and each \gls{ue} is simultaneously served by all \glspl{ap}, with \(M > K\). 
The channel between \gls{ap} \(m\) and \gls{ue} \(k\) is
\begin{equation}
g_{m,k} = \sqrt{\beta_{m,k}} h_{m,k}
\end{equation}
where \(\beta_{m,k}\) denotes the large-scale fading component capturing the path loss and shadowing, and \(h_{m,k}\) represents the small-scale fading. By stacking all channel coefficients, the overall channel matrix is given by \(\mathbf{H} = [\mathbf{g}_1, \mathbf{g}_2, \ldots, \mathbf{g}_K] \in \mathbb{C}^{M \times K}\), where each column \(\mathbf{g}_k \in \mathbb{C}^{M}\) corresponds to the channel vector between all \glspl{ap} and \gls{ue} \(k\).

To jointly serve all users, linear precoding is employed across the \glspl{ap}. 
Let \(\mb{W}=\left[\mb{w}_1, \mb{w}_2, \ldots, \mb{w}_K\right] \in \mathbb{C}^{M \times K}\) denote the precoding matrix, where \(\mb{w}_k \in \mathbb{C}^{M}\) corresponds to the precoder targeting \gls{ue} \(k\). 
Accordingly, the received signal at \gls{ue} \(k\) can be expressed as
\begin{equation}
y_k = \mb{g}_k^\mathrm{T} \mb{w}_k s_k + \sum_{l=1,\, l\neq k}^K \mb{g}_k^\mathrm{T} \mb{w}_l s_l + n_k
\end{equation}
where \( \mb{g}_k \in \mathbb{C}^{M} \) denotes the channel vector between \gls{ue} \(k\) and all \glspl{ap}, and \( n_k \sim \mathcal{CN}(0, \sigma^2) \) represents additive white Gaussian noise. 
The transmitted symbols \( s_k \sim \mathcal{CN}(0,1) \) are independent and identically distributed (i.i.d.) across all users. Based on this model, the received \gls{sinr} at \gls{ue} \(k\) is given by
\begin{equation}
\operatorname{SINR}_k = \frac{\left|\mb{g}_k^{\mathrm{T}} \mb{w}_k\right|^2}
{\sum_{l=1,\, l \neq k}^K \left|\mb{g}_k^{\mathrm{T}} \mb{w}_l\right|^2 + \sigma^2}.
\end{equation}
The overall system throughput/sum rate is
\begin{equation}
R_{\text{sum}} = \sum_{k=1}^{K} R_k 
= \sum_{k=1}^{K} \log_2\left(1 + \operatorname{SINR}_k\right).
\label{equ:1}
\end{equation}
The sum rate in~\cref{equ:1} serves as the \gls{kpi} for evaluating the effectiveness of the precoding strategy. 

\subsection{Precoding Schemes}
Neural networks with sufficient neurons can approximate any continuous non-linear function, implying their potential to learn the mapping from the channel matrix \(\mathbf{H}\) to the precoding matrix \(\mathbf{W}\)~\cite{ref15}. 
A straightforward candidate is the \gls{mlp}, but its enormous hypothesis space leads to excessive parameters, high data demand, and prohibitive inference complexity.
A more efficient approach is to introduce architectural inductive biases that constrain the hypothesis space. In \gls{dmimo}, the \gls{ap}--\gls{ue} connectivity naturally forms a bipartite graph, enabling \glspl{gnn} to exploit this structure through message passing. Their inherent \emph{permutation equivariance} ensures that permuting users or antennas only permutes the corresponding precoder entries, thereby reducing complexity while maintaining expressiveness~\cite{ref27}.
Such graph-based inductive biases are particularly valuable when dealing with hardware-induced nonlinearities, where linear compensation methods (e.g., \gls{z3ro}~\cite{ref25}) become ineffective for higher-order distortions~\cite{ref24}.
Consequently, \gls{gnn}-based precoding is selected as our primary learning-based scheme, whose detailed structure and deployment will be presented in~\cref{gnn}. 

For comparison, we also consider classical linear precoders. 
Specifically, \gls{cb}, corresponding to \gls{mrt}, aligns each precoding vector with the conjugate of the channel, while \gls{rzf} suppresses multi-user interference by inverting the composite channel matrix as~\cite{ref2}
\begin{equation}
    \mathbf{W}= 
    \begin{cases}
        \mb{H}^H & \text{for CB} \\ 
        \mb{H}^H \left( \mb{H}\mb{H}^H + \alpha \mb{I} \right)^{-1} & \text{for RZF}
    \end{cases}
\end{equation}

It is worth noting that the objective of employing the \gls{gnn}-based precoding scheme in this work is not to outperform conventional linear methods such as \gls{cb} or \gls{rzf}, but rather to evaluate the implementation feasibility and performance of \gls{ai}-based precoders within a practical testbed prototype. 

\section{System Implementation}
\label{sec:system implemetation}
To validate the \gls{gnn}-based precoding scheme in practice, we implemented it in the \gls{dmimo} testbed Techtile~\cite{ref26}. This section first introduces the hardware setup and synchronization procedures, and then describes the deployment of the \gls{gnn}-based precoder within the transmission protocol.

\subsection{Testbed Set-Up}
\begin{figure*}[t]
    \centering
    \begin{subfigure}[t]{0.55\textwidth}
    \centering
    \includegraphics[height=2in]{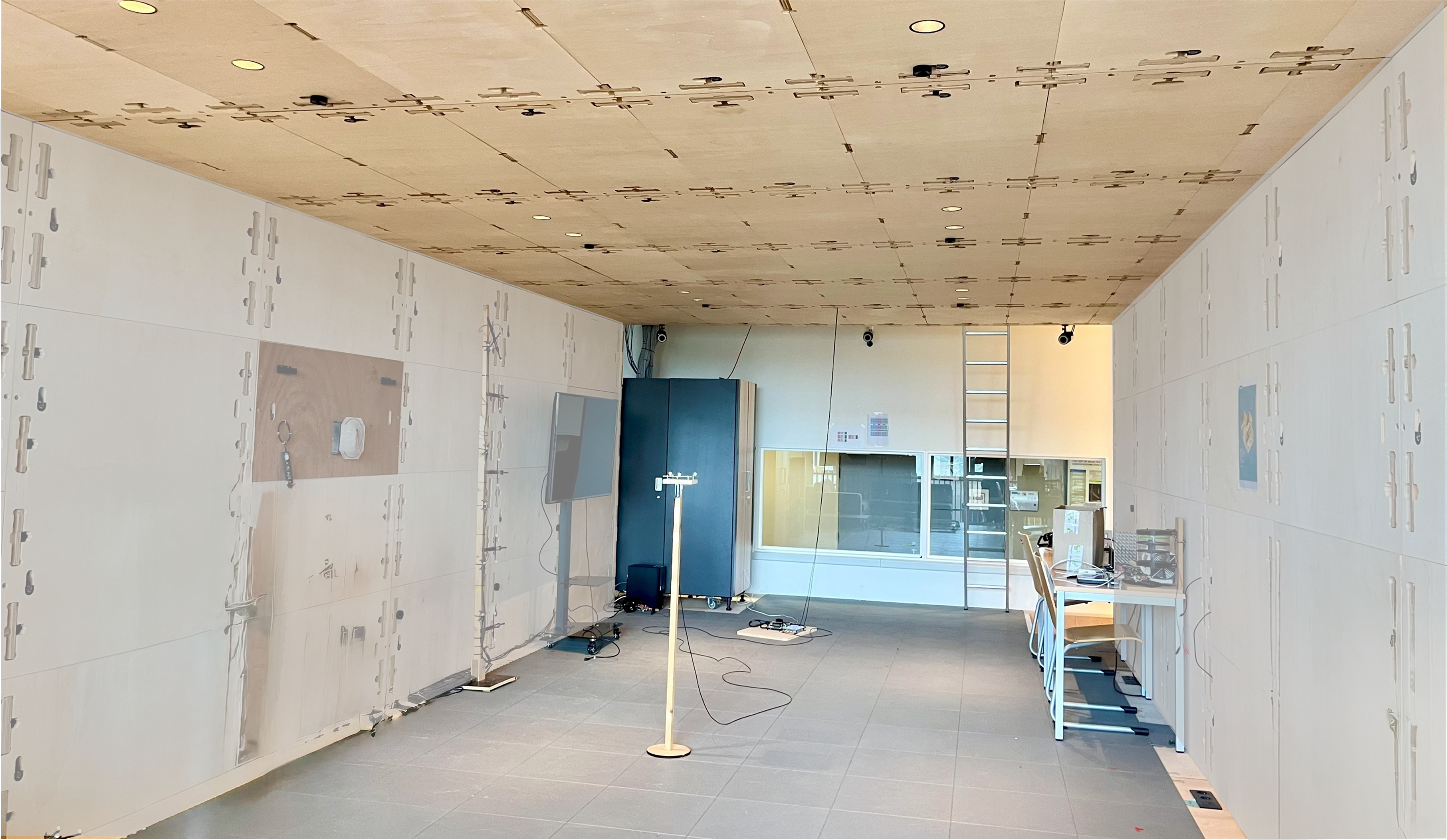}
    \end{subfigure}%
    \qquad
    \begin{subfigure}[t]{0.25\textwidth}
    \centering
    \includegraphics[height=2in]{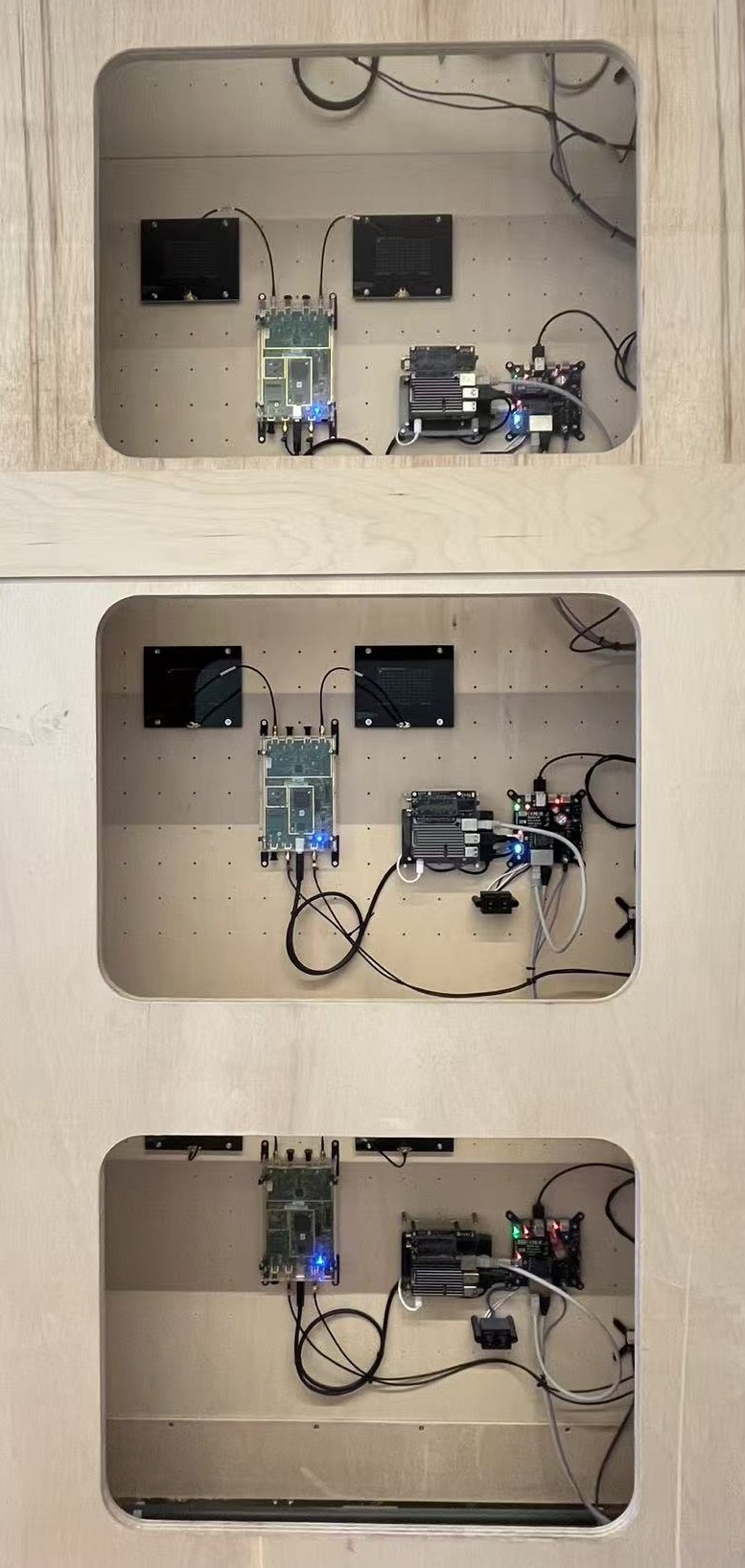}%
    \end{subfigure}
    \caption{\small Left: The Techtile support structure hosting 140 tiles -- Right: The back of three of such tiles, equipped with the default setup, i.e., a \gls{sdr} (\gls{usrp} B210), processing unit (\gls{rpi} 4) and power supply with Power-over-Ethernet. Each tile is connected to the central unit via a single Ethernet cable, which carries both power and data.}\label{fig:picture-techtile}
\end{figure*}

The experimental setup is conducted in Techtile, shown in~\cref{fig:picture-techtile}. Techtile is a distributed \gls{sdr}-based infrastructure where each tile contains one \gls{usrp} NI B210 device~\cite{ref26}. A single \gls{ue} is used in this work.
Operating in full-duplex mode, the B210 provides access to four RF channels and supports a maximum transmit power of \SI{20}{\dBm}. 
In our setup, we deployed up to \num{33} \glspl{usrp} in the ceiling, acting as distributed \glspl{ap} for downlink transmission.
All devices operate at a carrier frequency of \SI{920}{MHz} with a sampling rate of \SI{250}{kHz}. 
A \gls{tdd} frame structure is adopted to exploit channel reciprocity. Prior to downlink beamforming, the setup is calibrated as shown in in~\cref{fig:tdd-flow}.
To support time and frequency synchronization, a clock distribution module (NI OctoClock CDA-2990) delivers reference signals through a \SI{10}{MHz} frequency source and a \gls{pps} time signal. 
They are synchronized by a grandmaster clock, which itself is aligned using \gls{gnss}. 

In practical implementation, the observed phase at the receiver is affected not only by the wireless channel but also by the transmit and receive \glspl{lo}, hardware distortions and transmission line delays~\cite{ref14}. 
To phase calibrate all \glspl{ap}, a \SI{920}{\mega\hertz} reference is distributed over a coaxal cable to all \glspl{ap}. Three phase measurements are performed at each \gls{ap}~\(i\). First, the reference phase \(\phi^{\text{Ref}}_i = \phi^{\text{ref}} - \phi^{\text{RX}}_i + \phi^{\text{cable}}_i\) explicitly includes the known cable delay term \(\phi^{\text{cable}}_i\), which is pre-measured and fixed for each connection.
Next, the pilot phase \(\phi^{\text{Pilot}}_i = \phi^{\text{pilot}} - \phi^{\text{RX}}_i + \phi^{\text{ch}}_i\) captures the channel-induced phase \(\phi^{\text{ch}}_i\). 
Finally, the loopback phase \(\phi^{\text{Loop}}_i = \phi^{\text{TX}}_i - \phi^{\text{RX}}_i\) represents the transmitter-side hardware phase ambiguity. 
The term \(\phi^{\text{RX}}_i\) accounts for the aggregate phase of the \(i\)-th RX chain, including both the propagation delay along the physical path and the receiver hardware-induced phase offset.

\paragraph{Uplink phase alignment}
During uplink pilot transmission, each AP observes the received pilot phase \(\phi^{\text{Pilot}}_i\), 
which does not contain the cable delay present in the reference measurement \(\phi^{\text{Ref}}_i\). 
To ensure consistency with the reference signal and correctly estimate the effective channel phase, 
the cable contribution must therefore be added to the pilot-based estimate
\begin{equation}
\phi^{\text{CSI}}_i = \phi^{\text{Pilot}}_i + \phi^{\text{cable}}_i.
\label{eq:phi_csi}
\end{equation}
This adjustment aligns the phase reference of the estimated \gls{csi} with that of the calibration measurement, 
ensuring that all \glspl{ap} share a common phase reference for downlink precoding.

\paragraph{Downlink phase compensation}
Based on the effective CSI obtained in~\cref{eq:phi_csi}, the precoder generates beamforming coefficients \(\theta^{\text{bf}}_i\) to achieve coherent signal combination at the \gls{ue}. 
During transmission, additional phase distortions arise from the transmit hardware and the same cable delay that affected the uplink. 
To preserve the intended precoding phase at the air interface, these distortions are compensated prior to transmission by subtracting the loopback- and cable-induced offsets as \(\theta^{\text{tx}}_i = \theta^{\text{bf}}_i - \phi^{\text{Loop}}_i - \phi^{\text{cable}}_i\), ensuring that the radiated signals are phase-aligned at the \gls{ue} and follow the desired precoding strategy.

\subsection{GNN-based Precoding Deployment}
\label{gnn}

% \begin{figure}[htbp]
%   \centering
%   \includestandalone[width=0.9\linewidth]{Figure/tdd_flow}
%   \caption{\Gls{tdd}-based deployment. The environment can be considered static as nothing is moving during the capture time, yielding a long coherence time.}
%   \label{fig:tdd-flow}
% \end{figure}
\begin{figure}[htpb]
\centering
\includegraphics[width=0.45\textwidth]{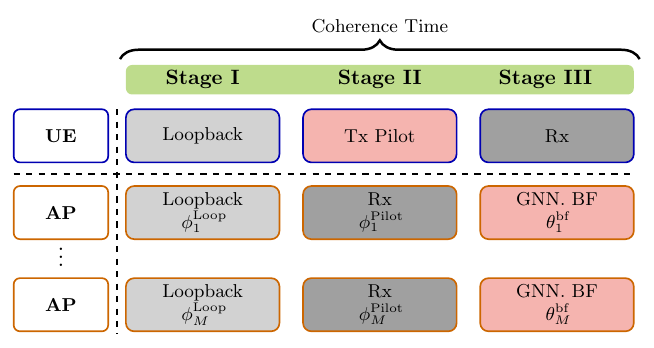}
\caption{\Gls{tdd}-based deployment. The environment can be considered static as nothing is moving during the capture time, yielding a long coherence time.}
\label{fig:tdd-flow}
\end{figure}

The deployed \gls{gnn}-based precoder learns the mapping from the channel matrix~\(\mb{H}\) to the precoding matrix~\(\mb{W}\) in an unsupervised manner, directly maximizing the sum-rate objective in~\cref{equ:1}. The network consists of eight stacked edge-centric message-passing layers operating on the bipartite \gls{ap}--\gls{ue} graph, where each edge represents a wireless link and its associated \gls{csi} serves as input features. Unlike node-centric designs, the embeddings are maintained on the edges and iteratively updated by aggregating contextual information from their incident nodes using a permutation-invariant mean operator, followed by a LeakyReLU activation. At the final layer, each edge embedding is mapped to the real and imaginary parts of the corresponding precoding coefficient, and the resulting precoding matrix is normalized to satisfy the total transmit power constraint~\cite{ref19}. The model is trained offline and then integrated at the \gls{cpu} side for inference, ensuring that it is fully trained and validated prior to deployment.

To validate the proposed framework on hardware, a \gls{tdd} protocol is designed, as illustrated in~\cref{fig:tdd-flow}. The entire procedure is executed within one channel coherence interval. In Stage~I, both the \gls{ue} and the \glspl{ap} perform loopback calibration to measure the internal phase offsets of their transceiver chains. In Stage~II, the \gls{ue} antenna transmits a continuous-wave pilot signal, i.e., an unmodulated sine tone at the carrier frequency, which serves solely as the pilot for subsequent processing. The received pilots are processed at the distributed units and forwarded to the \gls{cpu}, which computes the aggregated \gls{csi}. In Stage~III, the estimated channel matrix is processed by the deployed \gls{gnn} to perform online inference and generate the precoding matrix within the same channel coherence interval as the received pilot. The result is then broadcast to all \glspl{ap} for downlink transmission.

\section{Results}
\label{sec:results}
\subsection{Fine-Tuning with Real Channel Measurements}
To evaluate the network performance using synthetic data, the large-scale fading is modeled according to the \gls{3gpp} Indoor Hotspot (InH) \gls{nlos} scenario~\cite{ref3}, as
\begin{equation}
\beta_{m,k} = 32.4 + 31.9\log_{10}(d_{m,k}) + 20\log_{10}(f_c),
\label{pathloss}
\end{equation}
where \(d_{m,k}\) denotes the distance (in meters) between \gls{ap}~\(m\) and \gls{ue}~\(k\), and \(f_c\) is the carrier frequency (in~GHz). 
The small-scale fading coefficients are assumed to be \gls{iid}, following \(h_{m,k} \sim \mathcal{CN}(0,1)\), indicating that each coefficient follows a complex Gaussian distribution.
The \gls{gnn}-based precoder is pretrained in an unsupervised manner on \num{100000} synthetic channel realizations for \num{150} epochs with a learning rate of \(10^{-4}\), batch size \num{256}, and \num{10000} samples reserved for validation and testing, using the Adam optimizer~\cite{ref30}.
% \begin{figure}[htbp]
%   \centering
%   \includestandalone[width=0.8\linewidth]{Figure/pretrfin}
%   \caption{Sum-rate performance of the \gls{dmimo} system with different numbers of \glspl{ue} (\(M=33\)) and precoding schemes. All methods are evaluated on real-world \gls{csi} data.}
%   \label{fig:network_perfor}
% \end{figure}
\begin{figure}[htpb]
\centering
\includegraphics[width=0.45\textwidth]{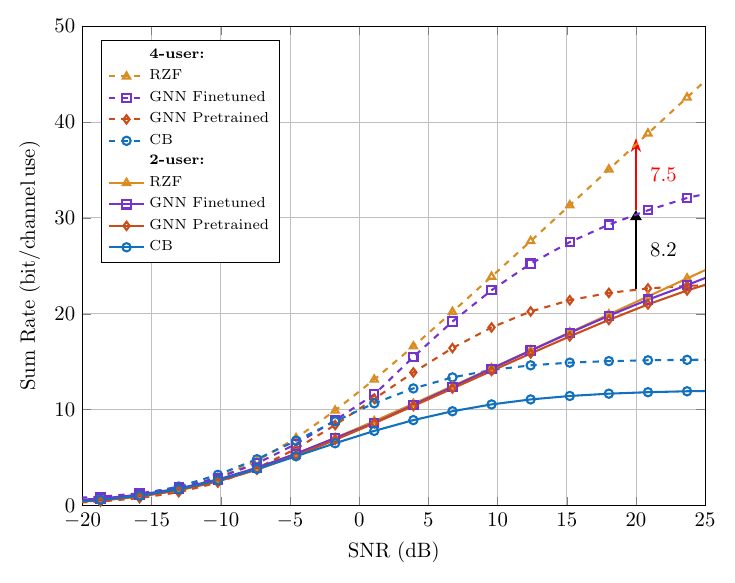}
\caption{Sum-rate performance of the \gls{dmimo} system with different numbers of \glspl{ue} (\(M=33\)) and precoding schemes. All methods are evaluated on real-world \gls{csi} data.}
\label{fig:network_perfor}
\end{figure}

To adapt the pretrained model to real-world propagation conditions, we perform fine-tuning by partially freezing the pretrained \gls{gnn} layers as proposed in~\cite{ref19}, which enables efficient domain adaptation with limited real data. 
The fine-tuning dataset consists of \num{500} real channel measurements collected from \num{33} ceiling-mounted \glspl{ap} in the Techtile testbed. 
A \gls{ue} placed on the floor transmitted \SI{40}{mW} pilot signals, and the receiver position was varied to ensure spatial diversity. 
Multi-user channel realizations were generated following the pairing strategy in~\cite{ref19} to emulate realistic communication scenarios. The dataset is publicly available, with access details provided on the first page.

As shown in~\cref{fig:network_perfor}, the pretrained \gls{gnn} surpasses \gls{cb} and approaches the performance of \gls{rzf} in the two-user case, while a larger gap appears for four users. 
After fine-tuning with real \gls{csi}, this gap is reduced from \SI{15.7}{bits/channel\,use} to \SI{7.5}{bits/channel\,use} (\SI{15.7}{\percent} improvement), demonstrating effective adaptation to practical interference conditions. 
Compared with a baseline network trained from scratch on real data, the fine-tuned model achieves higher sum-rate due to better initialization and domain adaptation under limited dataset size.

\subsection{Extrapolation and Interpolation Evaluation}
\begin{figure}[htpb]
\centering
\includegraphics[width=1\linewidth]{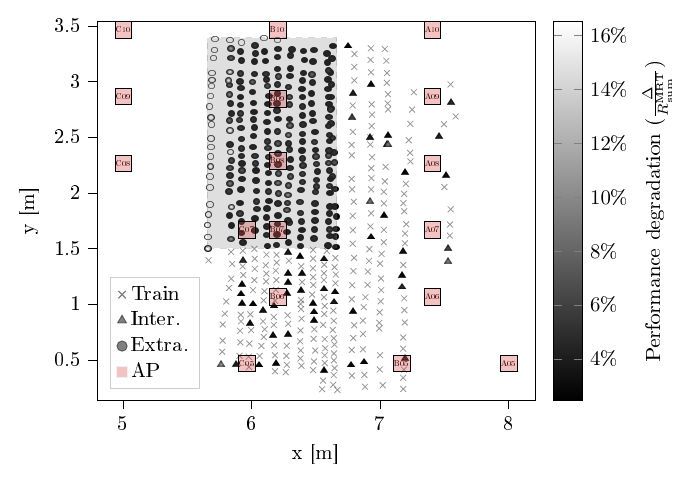}
\vspace{-20pt}
\caption{Heatmap of extrapolation and interpolation performance gap (\(\Delta = R_{\text{sum}}^{\text{MRT}} - R_{\text{sum}}^{\text{GNN}}\)).}
\label{fig:point-wise-heatmap}
\end{figure}

To assess the generalization of the fine-tuned \gls{gnn}-based precoder, we conduct point-wise validation using real channel measurements across two spatial scenarios: interpolation (test samples within the training region) and extrapolation (samples from unseen regions).
As shown in Fig.~\ref{fig:point-wise-heatmap}, a subset of the measurement area is masked to form the extrapolation region, while the rest is used for training and validation.
The rate difference between the learned precoder and \gls{mrt} benchmark is visualized as a heatmap to reveal spatial consistency.

% \begin{figure}[htbp]
%   \centering
%   \includestandalone[width=0.9\linewidth]{Figure/training_sample_efficiency}
%     \caption{Sample efficiency (shaded regions show variability across random seeds, not the full distribution support).}
%   \label{fig:Sample efficiency}
% \end{figure}
\begin{figure}[htpb]
\centering
\includegraphics[width=0.45\textwidth]{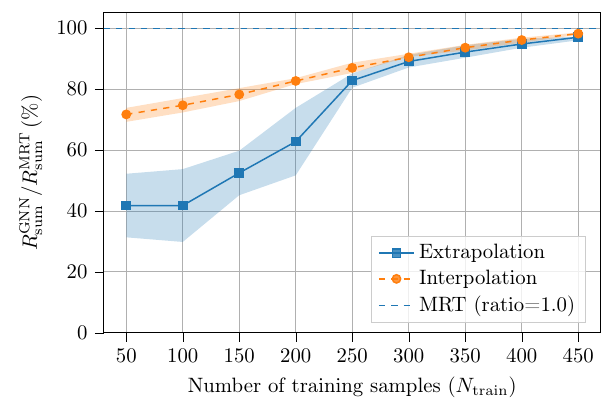}
\caption{Sample efficiency (shaded regions show variability across random seeds, not the full distribution support).}
\label{fig:Sample efficiency}
\end{figure}
The \gls{gnn}-based precoder achieves near-\gls{mrt} performance in the interpolation region, demonstrating strong spatial consistency where training samples are available. 
However, in the extrapolation region, where the test points are spatially separated from the training set, the achievable rate degrades noticeably, with an average loss of up to \SI{0.7}{bits/channel\,use} compared with the \gls{mrt} benchmark. 
This degradation confirms that while fine-tuning on real data improves adaptation to practical propagation conditions, it also limits spatial generalization beyond the measured domain. 
These results highlight the importance of collecting diverse and spatially distributed \gls{csi} samples to enhance model robustness under realistic deployment scenarios.

To further assess the data efficiency of real-world measurements in fine-tuning the \gls{gnn}-based precoder,~\cref{fig:Sample efficiency} illustrates the relationship between the number of training samples \(N_{\text{train}}\) and the ratio between the fine-tuned \gls{gnn} performance and the \gls{mrt} benchmark in both interpolation and extrapolation settings. 
As \(N_{\text{train}}\) increases, the performance ratio steadily approaches unity, indicating that additional real-world samples enable the model to better adapt to practical propagation conditions. 
Although the extrapolation performance initially lags behind the interpolation case, both eventually converge toward the \gls{mrt} baseline, demonstrating the model's improved robustness with increased training data. Moreover, the variance across random seeds is notably smaller in the interpolation regime, suggesting that the model achieves more stable convergence when trained on in-distribution samples.

\subsection{End-to-End Downlink Transmission Validation}

\begin{figure*}[t]
\centering
\begin{subfigure}{1\textwidth}
\includegraphics[width=1\linewidth]{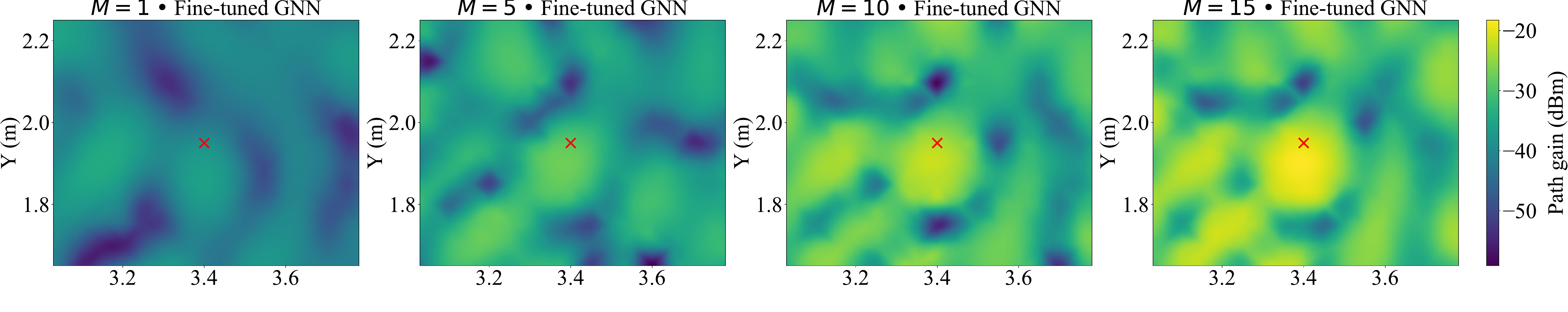}
% \caption{The heatmap obatained in the target area with \gls{gnn}-based precoder deployment.}
\label{fig:4Theatmap_gnn}
\end{subfigure}
\begin{subfigure}{1\textwidth}
\vspace{-15pt}
\includegraphics[width=1\linewidth]{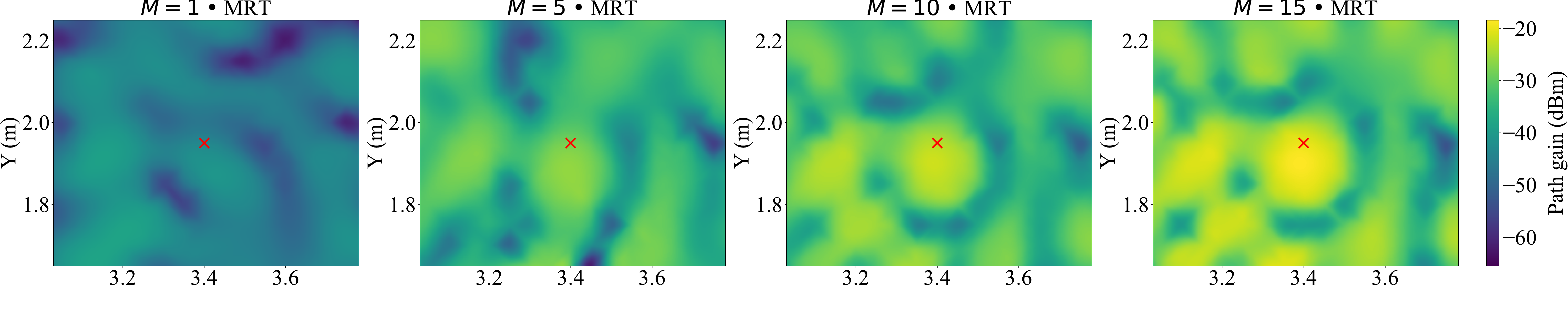}
\label{fig:4Theatmap_mrt}
\end{subfigure}
\begin{subfigure}{1\textwidth}
\vspace{-15pt}
\includegraphics[width=1\linewidth]{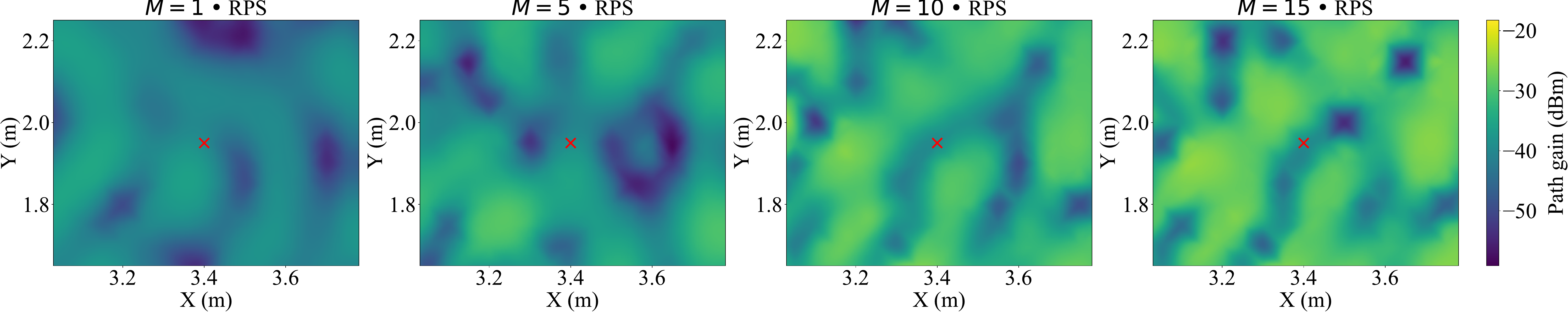}
% \caption{The heatmap obtained in the target area with \gls{rps}-based precoder deployment.}
\label{fig:4Theatmap_rps}
\end{subfigure}
\vspace{-20pt}
\caption{Received power heatmaps in the target area under different precoding schemes and varying numbers of cooperating \glspl{ap}, each transmitting with equal power. 
The first, second, and third rows correspond to the fine-tuned \gls{gnn}-based precoder, the \gls{mrt} benchmark, and the \gls{rps} scheme, respectively. The \gls{ue} is located at \((3.4, 1.95)\) in all scenarios.}
\label{fig:4Theatmap}
\end{figure*}

In this subsection, we experimentally validate the end-to-end downlink performance of the fine-tuned \gls{gnn}-based precoder in the single-user case on the Techtile testbed, and compare it with benchmark schemes following~\cref{gnn}.  
All \glspl{ap} transmit with an equal power of \SI{0}{\dBm}.
The heatmaps in~\cref{fig:4Theatmap_gnn} illustrate the received power distribution within the target area under different numbers of cooperating \glspl{ap}.  

In the \gls{siso} case, where only one \gls{ap} transmits, no apparent power focus is observed at the \gls{ue} due to the absence of coherent combining.  
As more \glspl{ap} participate, the received power increasingly concentrates around the \gls{ue}, reflecting enhanced phase alignment among distributed transmitters.  
For example, with \(M=5\) and \(M=10\), distinct power peaks emerge at the target position, while for \(M=15\), the focal spot becomes sharper and stronger.  
The fine-tuned \gls{gnn}-based precoder achieves received power levels comparable to the \gls{mrt} benchmark across all settings, demonstrating near-optimal coherent combining.  
In particular, the \(M=15\) setup yields a received power of \SI{-19.5}{\dBm}, corresponding to a \SI{20.9}{dB} gain over the \gls{siso} case, close to the theoretical \(20\log(M)\) scaling. 
Note that this theoretical gain applies to colocated \gls{mimo} systems; in distributed settings with varying large-scale fading, the actual gain may deviate from this ideal trend. 

For comparison, the \gls{rps} scheme, which randomly changes the transmit phases of all antennas at regular intervals while maintaining time and frequency synchronization, is also evaluated.  
Under this non-coherent transmission strategy, the received power increases from \SI{-40.4}{\dBm} in the \gls{siso} case to \SI{-30.6}{\dBm} with \(15\) transmitters, yielding an \SI{9.8}{dB} gain that closely matches the theoretical \(10\log(M)\) non-coherent combining expectation~\cite{ref20}.  
This contrast highlights the effectiveness of the proposed hardware synchronization and calibration scheme in enabling coherent combining across distributed transmitters.

\section{Conclusions}
\label{sec:conclusions}
Experimental results showed a \SI{15.7}{\percent} performance gain in the multi-user case after fine-tuning, while in the single-user scenario the model achieved near-\gls{mrt} performance with less than \SI{0.7}{bits/channel,use} degradation on unseen positions.
Data efficiency analysis further revealed consistent performance gains as the number of real-world training samples increased.
End-to-end validation on the reciprocity-calibrated testbed confirmed coherent power focusing with a received power of \SI{-19.5}{\dBm} (about \SI{20.9}{dB} gain over \gls{siso}), demonstrating near-optimal combining.
These results verify the feasibility and robustness of learning-based precoding under realistic conditions and motivate future work toward the practical deployment of multi-user interference mitigation.
\printbibliography

\end{document}